\documentstyle[aps,prbbib,twocolumn,epsf]{revtex}

\begin{document}
\draft
\title{Weakly Nonlinear AC Response: Theory and Application}

\author{Zhong-shui Ma$^{1,2}$, Jian Wang $^1$ and Hong Guo$^{1,3}$}
\address{1. Department of Physics, The University of Hong Kong, 
Pokfulam Road, Hong Kong, China\\
2. Advanced Research Center, Zhongshan University, Guangzhou, China\\
3. Center for the Physics of Materials and Department 
of Physics, McGill University, Montreal, PQ, Canada H3A 2T8\\}
\maketitle

\begin{abstract}
{\bf 
We report a microscopic and general theoretical formalism for 
electrical response which is appropriate for both DC and AC 
weakly nonlinear quantum transport. The formalism emphasizes the
electron-electron interaction and maintains current conservation
and gauge invariance. It makes a formal connection between linear 
response and scattering matrix theory at the weakly nonlinear level. 
We derive the dynamic conductance and predict the 
nonlinear-nonequilibrium charge distribution. The definition of
a nonlinear capacitance leads to a remarkable scaling relation
which can be measured to give microscopic information about a conductor.
}
\end{abstract}

\pacs{73.23.Ad,73.23.Ps,72.10.Bg}
\medskip

Over the last decade tremendous effort have been devoted in developing 
quantum transport theories which are appropriate for quantum coherent 
conductors\cite{datta}. While substantial progress have been achieved, 
so far there is still a lack of a general formalism which works not only 
in the linear DC regime but also in both DC and AC nonlinear regimes.
Clearly, due to the great importance to technological applications 
of quantum conductors, such a formalism is urgently in need. AC
transport involves time dependent fields, thus induction is 
important as characterized by, {\it e.g.} the existence of
displacement current. Many transport theories do not consider this 
ingredient, resulting to a violation of current conservation as 
observed by B\"uttiker\cite{buttiker1}. The nonlinear DC transport 
coefficients appear in front of powers of external bias voltage in
the expression of electric current: $I=I(\{V_\alpha\})$ where 
$V_{\alpha}$ is the potential at a probe labeled $\alpha$
which connects to the electron reservoir with chemical potential
$\mu_\alpha$. Thus a correct theory must maintain gauge invariance
in addition to the current conservation: the physics should not change 
when potential everywhere is shifted by the same constant amount. This 
has severe constraints on the nonlinear DC transport 
coefficients\cite{buttiker1}.

So far current-conserving and gauge invariant quantum transport 
has been analyzed using the scattering matrix theory (SMT)
as developed by B\"uttiker and co-workers\cite{buttiker1,buttiker3}.
Within this formalism, by including electron-electron (e-e) interaction, 
current conservation is obtained up to linear order of AC frequency 
$\omega$ for the dynamic conductance $G_{\alpha\beta}(\omega)$, and for 
special cases to the $\omega^2$ term. The same interaction has 
led to a gauge invariant second order weakly nonlinear DC conductance.
In SMT\cite{buttiker1}, various $\omega$-independent partial density of 
states (PDOS) appear naturally which characterizes the scattering. The 
SMT is quite intuitive and can be implemented numerically for practical 
calculations\cite{wang1}. However, so far no theoretical formalism
exists which goes beyond the linear-$\omega$ AC and second order DC
transport coefficients, and which satisfies current conservation and
gauge invariance. There is also no approaches available to analyze the
weakly nonlinear AC transport. Hence up to now these important problems 
have not been investigated systematically.

It is well known that the {\it linear} coherent quantum transport can 
be discussed using either scattering matrix or using linear 
response\cite{fisher,baranger}. In the linear DC regime, connection of 
these two approaches has been formalized\cite{baranger} and attempt has 
been made to generalize it to nonlinear DC situation\cite{vegvar}. 
Since a response theory is usually easier to be generalizable to 
higher order as it is in a perturbative form\cite{callaway}, we thus 
expect a general theoretical formalism derivable from the response 
theory which is appropriate for both DC and AC weakly {\it nonlinear} 
regimes in addition to {\it linear} situations. We have indeed 
achieved this goal and it is the purpose of this Letter to 
report this development.

Before go into the details, we summarize the main results:
(1) A general formalism for deriving AC and DC transport 
coefficients has been found which is current conserving and gauge 
invariant. (2) The formalism is put into a form which is 
numerically calculable for mesoscopic or even microscopic
conductors order by order in weak bias and to all orders in 
$\omega$. (3) The results obtained from SMT are contained in this
formalism thus we make the formal connection between SMT and the
response theory at the weakly nonlinear DC and AC level.
(4) Generalized notions of the {\it characteristic potential}, 
the {\it nonequilibrium charge}, and the Lindhard functions are derived 
naturally from the time dependent internal response. (5) As examples of 
this formalism, we derived the linear (in bias) and second order weakly 
nonlinear dynamic conductance to all orders of $\omega$. We 
also predict the behavior of a nonlinear ``capacitance'', 
and derive a remarkable scaling relation between the capacitances 
which is experimentally measurable and which gives 
microscopic information such as the DOS of a system.

A response theory considers a time dependent 
perturbation\cite{baranger} 
$H' \equiv V({\bf r},t)=V({\bf r})e^{-\delta |t|}cos(\Omega t)$
to the Hamiltonian $H_o$ of a conductor. Here $H_o$ contains the 
kinetic energy, the influence of an external static magnetic field, 
the confining potential of the conductor, or any other static
potential. We assume that the problem of $H_o$ has been solved, 
$H_o\psi_n = \epsilon_n\psi_n$, and we are interested in the effect of 
$H'$. It is crucial to realize that just solving the problem of 
$H_o + H'$ will not generate a current conserving and gauge invariant 
transport theory, since the time varying field induces an internal 
potential $U(\{V\},t)$ which is a functional of the external 
perturbation $V$. It is this internal potential which generates such 
effect as a displacement current. Hence one must solve a quantum 
mechanical problem in conjunction with the Maxwell equations. We thus 
start from a self-consistent (up to the Hartree level) Hamiltonian $H$ 
coupled with the Helmhotz equation:
\begin{equation}
H\ =\ H_o + H' (t) + U(\{V\},t)
\label{hh}
\end{equation}
\begin{equation}
\nabla^2U({\bf r},t)-{1\over c^2}{\partial^2 U({\bf r},t)
\over \partial t^2}=-4\pi\delta\rho ({\bf r}, t)\ ,
\label{helmhotz}
\end{equation}
where $\delta \rho$ is the total charge involved in the response
including those of the induced.

We shall be interested in a multi-probe conductor small enough such 
that quantum coherence is maintained. Inside a probe $\alpha$ far
away from the scattering region, the amplitude of the external
disturbance is written as $V_{\alpha}$. We shall calculate the internal
potential $U$ in a series form, expanded in powers of $V_{\alpha}$
using the {\it characteristic potentials} (CP) $u_\alpha({\bf r},t)$,
$u_{\alpha\beta}({\bf r},t)$ etc:
\begin{equation}
eU =\sum_\alpha u_\alpha V_\alpha
+{1\over 2}\sum_{\alpha\beta}u_{\alpha\beta}V_\alpha V_\beta
+\cdots
\label{eu}
\end{equation}
This expansion makes sense at weakly nonlinear regime. 
The CP characterizes the internal potential response to an external
time-dependent perturbation. The physics associated with weakly
nonlinear quantum transport is related to the CP's of the appropriate
order (see below). Gauge invariance puts constraints on the
CP, for instance $\sum_\alpha u_\alpha ({\bf r},t)=1$,
$\sum_\alpha u_{\alpha\beta}({\bf r},t)
=\sum_\beta u_{\alpha\beta} ({\bf r},t)=0$, and in general
$\sum_{\gamma\in\beta} u_{\alpha\{\beta\}_l}=0$. Here the subscript
$\{\beta\}_l$ is a short notation of $l$ indices
$\gamma,\delta,\eta,\cdot\cdot\cdot$. The quantum mechanics problem of 
Eq. (\ref{hh}) is solved in the standard series fashion by iterating the 
Liouville-von Neumann equation for the density operator\cite{callaway}
${\hat{\rho}}({\bf r},t)={\hat{\rho}}_0+
\sum_{l=1}\delta{\hat{\rho}}_l({\bf r},t)$.
Using the Liouville-von Neumann equation, order by order we thus
derive equations for the terms $\delta {\hat{\rho}}_l({\bf r},t)$. These
equations can be formally solved in frequency space in terms
of the characteristic potentials at the appropriate order:
$\delta{\hat{\rho}}_1$ is related to $u_\alpha$, 
$\delta{\hat{\rho}}_2$ is related to $u_{\alpha\beta}$ and 
$u_\alpha$, $\delta{\hat{\rho}}_3$ to $u_{\alpha\beta\gamma}$ and 
the lower order $u$'s, etc. Then using the density operator we can 
compute physical quantities such as the charge density distribution 
and the electric current\cite{callaway}.

With the charge density distribution and the Helmhotz 
equation (\ref{helmhotz}), we can derive equations for the 
characteristic potential which is one of the central results of 
this work:
\[-\nabla^2u_{\{\alpha\}_l}({\bf r},\omega)-
\frac{\omega^2}{c^2}u_{\{\alpha\}_l}({\bf r},\omega)\]
\[
+4\pi e^2\int d{\bf r}_1\Pi ({\bf r},{\bf r}_1,\omega)
u_{\{\alpha\}_l}({\bf r}_1,\omega)\] 
\begin{equation}
=\ 4\pi e^2 \frac{dn_{\{\alpha\}_l}({\bf r},\omega)}{d\epsilon}\ .
\label{result1}
\end{equation}
The right hand side of Eq. (\ref{result1}) is given by the frequency 
dependent local density of states (LDOS) which can be derived using the
Green's functions. For example the lowest order (one index) LDOS is
given by
\[\frac{dn_\alpha({\bf r},\omega)}{d\epsilon}=-\frac{\hbar^2}{2m}\times\]
\begin{equation}
\int dy_{1\alpha}\sum_{mn}\frac{f_{nm}}{\epsilon_{nm}}
\frac{\psi_n^*({\bf r})\psi_m({\bf r}){\bf W}_{mn}({\bf r}_1)\cdot
\hat{{\bf x}}_{1\alpha}}{\epsilon_{nm}+\hbar\omega+i\eta}
\label{dnde}
\end{equation}
where\cite{baranger}
${\bf W}_{nm}\equiv \psi^*_n({\bf r}){\stackrel{\leftrightarrow}
{\cal D}}\psi_m({\bf r})$ with 
${\stackrel{\leftrightarrow}{\cal D}}$ the standard double-sided 
derivative operator; $\epsilon_{nm}\equiv \epsilon_n -\epsilon_m$, 
$f_{nm}\equiv f(\epsilon_n)-f(\epsilon_m)$,
and $\eta$ is infinitesimal. The integration in (\ref{dnde}) is along the 
boundary of probe $\alpha$, and unit vector $\hat{{\bf x}}_{1\alpha}$ is
the direction along this probe. This expression can be further written
in terms of the Green's function. The third term on the left of Eq.
(\ref{result1}) is the induced charge written in terms of a frequency 
dependent Lindhard function which is given by
\begin{equation}
\Pi ({\bf r},{\bf r}_1,\omega)
={1\over e}\sum_{mn}\frac{f_{nm}
\psi^*_n({\bf r})\psi_m({\bf r})\psi^*_m({\bf r}_1)\psi_n({\bf r}_1)}
{\epsilon_{nm}+\hbar\omega+i\eta} .
\label{lindhard}
\end{equation}
Notice that the same Lindhard function appears in all equations. 
The boundary conditions can be set such that deep inside a reservoir 
connected to probe $\alpha$, where the e-e interaction is completely 
screened, $u_\alpha=1$ while all others equal to zero.

Some discussions of Eq. (\ref{result1}) are needed. (1) The characteristic 
potential has multiple indices which are necessary in order to study 
nonlinear (in bias) transport coefficients. (2) All the terms are 
$\omega$-dependent as required for a general theory which is appropriate
to all orders in $\omega$. It is the combination of multiple
indices and $\omega$-dependence of the CP which allows the analysis
of weakly {\it nonlinear} DC and AC transport in general terms
of $\omega$. (3) The Lindhard function as derived is calculable from 
single particle states. It recovers the static result 
of Ref. \onlinecite{levinson} by letting $\omega \rightarrow 0$. 
For a perfect one-dimensional ballistic wire it is consistent with 
that of Ref. \onlinecite{blanter} (including the double time derivative 
of Eq. (\ref{helmhotz})). (4) The concept of LDOS is generalized to higher 
order: at linear order (one index) it is the frequency dependent 
injectivity\cite{buttiker3}, while at higher order (more than one 
indices) it also contains some internal response of the lower order. 
While their expressions (not shown except Eq. (\ref{dnde})) are 
complicated, they are all expressed in terms of the Green's functions. 
(5) It can be shown that the Lindhard function is related to the 
first order (one index) total LDOS, 
$\int d{\bf r}_1\Pi ({\bf r},{\bf r}_1,\omega)
=dn({\bf r},\omega)/d\epsilon$, where 
$dn({\bf r},\omega)/d\epsilon=\sum_\alpha
dn_\alpha ({\bf r},\omega)/d\epsilon$.
 
After solving the characteristic potentials of order $l$
from Eq. (\ref{result1}), we thus obtain the total charge 
distribution at this order, $\delta\rho_l({\bf r},\omega)$, 
which is just the right hand side of (\ref{result1}) subtracting 
the term with the Lindhard function, multiplying the external 
potential variations $V_\alpha V_\beta V_\gamma \cdot\cdot\cdot$
(see Eq. (\ref{eu})). Various transport properties can be obtained 
immediately.

\noindent
\underline{{\bf Nonlinear ``Capacitance''.}} 
From the total charge distribution, we can write $Q_\alpha
=\sum_\beta C_{\alpha\beta}V_\beta + 1/2\sum_{\beta\gamma}
C_{\alpha\beta\gamma}V_\beta V_\gamma+\cdot\cdot\cdot$, where 
$Q_\alpha$ is just the appropriate spatial integration of the 
charge density. Hence the nonlinear theory naturally allows the 
definition of {\it nonlinear capacitance} coefficients,
\[C_{\{\alpha\}_{l}}(\omega )\ =\ 
\ e^2\int_\Gamma d{\bf r}
\frac{dn_{\{\alpha\}_{l-1}}({\bf r},\omega )}{d\epsilon}\]
\begin{equation}
-e^2\int_\Gamma d{\bf r}\int d{\bf r}_1
\Pi ({\bf r},{\bf r}_1,\omega)
u_{\{\alpha\}_{l-1}}({\bf r}_1,\omega )
\label{result2}
\end{equation}
where index $l>1$. The spatial integral over $\Gamma$ means 
integrating over the region where the charge $Q$ is positive 
(or negative). At linear order in both $\omega$ and voltage,
{\it i.e.} when $l=2$ and letting $\omega=0$, this gives the 
electrochemical capacitance\cite{buttiker3} which is of great 
experimental interest\cite{capa}. 
Note that due to the finite screening length resulting from small 
DOS for mesoscopic conductors, a nonequilibrium charge distribution 
can be established even when there is DC coupling between the two 
capacitor ``plates'' (the +Q and -Q regions)\cite{buttiker4}.
If we keep the general $\omega$ dependence but still work on 
the linear bias order ($l=2$), Eq. (\ref{result2}) gives the 
general linear dynamic response of the charge. A further very 
interesting result of (\ref{result2}) is the nonlinear 
``capacitance'' for $l>2$ and setting $\omega =0$. 
Such a quantity does not seem to have a geometrical counterpart, 
and it measures the degree of the nonequilibrium charge pile-up 
at the nonlinear order.

In particular, let's calculate $C_{\alpha\beta\gamma}$ for a 
parallel plate capacitor, where each plate has an area $A$ and 
is infinitely thin, and they are located in space on the $y-z$ plane 
at positions $x=0$ and $x=a$. Using Eq. (\ref{result1}) to solve 
the characteristic potentials at different regions from $x=-\infty$ to 
$x=+\infty$, matching the solutions at $x=0,a$, we obtain $u_1$ and
$u_{11}$. Within the Thomas-Fermi approximation of the Lindhard
function, from Eq. (\ref{result2}) we obtain
\begin{equation}
2e^4 \frac{C_{111}}{C_{11}^3}\ =\ \frac{d}{d\epsilon}
\left[ \left(\frac{dN_2}{d\epsilon}\right)^{-2}
      -\left(\frac{dN_1}{d\epsilon}\right)^{-2}\right]\ ,
\label{scaling}
\end{equation}
where $dN_i/d\epsilon$ is the total DOS on plate $i$, and 
$C_{11}=[4\pi a/A + \sum_{i=1}^2 1/(e^2dN_i/d\epsilon)]^{-1}$
is the usual electrochemical capacitance\cite{buttiker3}.
Eq. (\ref{scaling}) is a remarkable relation because on the left hand side 
there are only {\it macroscopic} quantities while on the right side
all are {\it microscopic}. We have checked that for another system,
being two large metallic rods with their ends at a distance $a$ 
apart, exactly the same scaling of $C_{111}/C_{11}^3$ is obtained,
the only difference being that the factor $2$ on the left side changes 
to 6. Hence this scaling relation gave us an extra handle on the
microscopics of a conductor: by measuring capacitances and forming
the scaling combination, all that is left is the energy derivatives
of DOS. For a symmetrical system Eq. (\ref{scaling}) gives $C_{111}=0$.
This is because $C_{111}$ is the coefficient of the term 
quadratic in bias, thus it must vanish for symmetric systems
because the value of the pile-up charge cannot change when 
$V_\alpha\rightarrow -V_\alpha$ for such a system.

\noindent
\underline{{\bf Linear Dynamic Conductance.}} As a second example
we derive the dynamic conductance $G_{\alpha\beta}(\omega)$.
$G_{\alpha\beta}$ is defined by the electric current flowing through the 
probe $\alpha$: 
${\bf I}_{\alpha}^{(1)}=\sum_\beta G_{\alpha\beta}(\omega) V_\beta$.
The current is obtained by a spatial integration of the current 
density across the transverse direction of the probe, with
the current density given by
${\bf J}_1({\bf r},\omega)=\sum_{nm}(\delta{\hat{\rho}}_1(\omega))_{nm}
({\bf J}_{op}({\bf r}))_{nm}$.  Here 
$({\bf J}_{op}({\bf r}))_{nm}=-i(e\hbar/2m){\bf W}_{nm}$.
With these definitions,
it is tedious but straightforward to evaluate $G_{\alpha\beta}(\omega)$. 
The final result can be cast into a very compact form {\bf exactly}:
\[G_{\alpha\beta}(\omega)=G_{\alpha\beta}(\omega =0)
-i\omega e^2\int d\epsilon (-f'(\epsilon))\left[
{dN_{\alpha\beta}(\omega)\over d\epsilon}\right.\]
\begin{equation}
\left.
-{1\over e}\int d{\bf r}_1d{\bf r}_2 {{d{\bar n}_\alpha ({\bf r}_1,
\omega)} \over d\epsilon} g({\bf r}_1,{\bf r}_2,\omega){{dn_\beta 
({\bf r}_2,\omega)}\over d\epsilon}\right]\ .
\label{result3}
\end{equation}
Here $G_{\alpha\beta}(\omega =0)$ is the familiar linear DC 
conductance which can be calculated using the transmission 
coefficient\cite{baranger}. $g({\bf r},{\bf r}_1,\omega)$
is Green's function of Eq. (\ref{result1}), 
$d{\bar n}_\alpha({\bf r}_1,\omega)/d\epsilon$ is another LDOS dual
of $dn_\alpha({\bf r}_1,\omega)/d\epsilon$, and 
$f'(\epsilon)=df/d\epsilon$. Setting $\omega=0$, this result reduces
exactly to the emittance obtained by SMT\cite{buttiker1,buttiker3}.
The $\omega$-dependent parts are given as a sum of two contributions. 
The first is due to the external perturbation at a reservoir, and 
it is determined by the $\omega$-dependent total PDOS
$dN_{\alpha\beta}(\omega)/d\epsilon$. The second is due to 
induction and is determined by the internal response. Although our
theoretical formalism can be proven to conserve current, this
is most easily seen from Eq. (\ref{result3}): it is not difficult
to verify $\sum_\alpha G_{\alpha\beta}(\omega)=0$. Finally
it can also be shown that the following Onsager relation holds
for the dynamic conductance:
$G_{\alpha\beta}(\omega, {\bf B})=G_{\beta\alpha}(\omega,-{\bf B})$.

\noindent
\underline{{\bf Weakly Nonlinear Dynamic Conductance.}}
Our theory can be used to analyze weakly nonlinear AC transport to
higher order in bias, and in the following we derive the second order
expression defined by the second order electric current
${\bf I}_{\alpha}^{(2)}=\sum_{\beta\gamma}
G_{\alpha\beta\gamma}(\omega) V_\beta V_\gamma$.
${\bf I}_{\alpha}^{(2)}$ is calculated in similar fashion as 
the ${\bf I}_\alpha^{(1)}$ of the last example, but using 
the density matrix in second order of bias. The final result is
\[G_{\alpha\beta\gamma}(\omega)=
G_{\alpha\beta\gamma}(\omega=0)
-i{e^2\over 2}\omega\left[\int d\epsilon 
(-f'(\epsilon)){dN'_{\alpha\beta\gamma}
(\omega)\over d\epsilon}\right. \]
\begin{equation}
\left. -{1\over e}\int d{\bf r}_1d{\bf r}_2
{d{\bar n}_\alpha({\bf r}_1,\omega)\over d\epsilon}
g({\bf r}_1,{\bf r}_2,\omega)
{dn_{\beta\gamma} ({\bf r}_2,\omega)\over d\epsilon}\right]\ .
\label{result4}
\end{equation}
In this result, $G_{\alpha\beta\gamma}(\omega=0)$ is the second order
weakly nonlinear DC conductance which was studied using
SMT\cite{buttiker1}. Hence setting $\omega=0$ (\ref{result4}) 
reduces to the known SMT result.
$\sum_{\alpha}dN'_{\alpha\beta\gamma}/d\epsilon$ 
equals the spatial integration of the second order nonlinear 
LDOS\cite{foot1} $dn_{\beta\gamma}({\bf r},\omega)/d\epsilon$.
Result Eq. (\ref{result4}) is general in frequency $\omega$. 
The current conservation and gauge invariance can be
explicitly confirmed: 
$\sum_\gamma G_{\alpha\beta\gamma}(\omega)=0$;
$\sum_\alpha G_{\alpha\beta\gamma}(\omega)
=\sum_\beta G_{\alpha\beta\gamma}(\omega)=0$.
In addition, we have checked that there seems no simple relationship
between $G_{\alpha\beta\gamma}(\omega,{\bf B})$
and $G_{\alpha\beta\gamma}(\omega,-{\bf B})$, in agreement with
the previous conclusion on this point\cite{vegvar}. 

Keeping terms linear in $\omega$, from Eq. (\ref{result4}) we derive 
the formula for the problem of second order Harmonic generation examined in 
Ref. \onlinecite{vegvar} (which was done at $\omega =0$). For the
example of double-barrier resonant tunneling device, near a resonance
energy $E\sim E_r$, the scattering matrix takes the Breit-Wigner form.  
This allows simple expressions for the CPs: 
$u_\alpha = \Gamma_\alpha/\Gamma$,
$u_{11}=-2e^2(\Gamma_1\Gamma_2/\Gamma^2)(\delta\epsilon/|\Delta|^2)$, 
where $\delta\epsilon\equiv (E-E_r)$, $\Delta\equiv \delta\epsilon +
i(\Gamma/2)$, $\Gamma_\alpha$ is the decay width due to barrier
$\alpha$, and $\Gamma=\Gamma_1+\Gamma_2$.  Evaluating all the terms of
Eq. (\ref{result4}) using these expressions, we derive
\begin{equation}
\frac{G_{111}(\omega)}{G_{111}(\omega=0)}\ =\
i\hbar\omega\frac{\Gamma}{2}\left(\frac{1}{|\Delta|^2}
-\frac{1}{\Gamma^2}\right)\ \ ,
\label{result5}
\end{equation}
where $G_{111}(\omega=0)$ is the second order weakly nonlinear DC
conductance\cite{buttiker1}. This result is also very interesting: it
suggests that by measuring the ratio of the nonlinear conductances one
obtains the microscopics of a tunneling device.

The above three examples demonstrate the power of the present
theoretical formalism: it is suitable for analyzing weakly
nonlinear DC and AC transport of mesoscopic conductors
systematically. The expressions (\ref{result3}) and (\ref{result4}) 
are general to all orders of $\omega$. Setting 
$\omega=0$ they recover exactly those of SMT\cite{buttiker1},
thus providing the connection between SMT and the response theory at 
the weakly nonlinear AC level.
In our formalism, the multiple indexed characteristic potential 
describes the variation of the scattering potential landscape of the 
conductor, at the appropriate weakly nonlinear level, as an external 
time dependent perturbation is applied to the electron reservoirs. 
The multiple indexed LDOS describes the induced charge at the weakly 
nonlinear level, and they are related to the scattering
Green's functions. For a mesoscopic conductor\cite{datta} 
or even an atomic scale conductor\cite{wan1}, the scattering Green's
function can be evaluated numerically thus our theoretical formalism 
provides the basis of numerical analysis for a variety of quantum 
conductors. Finally, we comment that in this paper we have used the 
Lorentz gauge for electrodynamics, we have made an explicit check that using
the Coulomb gauge gives exactly the same result. Clearly any gauge
should work as is required by the Maxwell equations.

Central to the theory presented here is the self-consistent coupling 
of the quantum mechanical equation with the Maxwell equations, thus
the internal response of the system is taking into account which is 
crucial to obtain electric current conservation and gauge invariance. 
Conceptually, we
have introduced the multiple indexed, frequency dependent 
characteristic potential and the local density of states. We have 
also extended the concept of nonequilibrium charge distribution
to the nonlinear order.
Our response theory is 
appropriate to equilibrium or near-equilibrium properties. For
far-from-equilibrium problems one may employ the nonequilibrium Green's 
functions\cite{wingreen}. A simple application of our theory naturally 
leads to the concept of nonlinear capacitance. A remarkable scaling 
relation is predicted such that a macroscopically measurable ratio of 
these capacitances is related to the microscopic information of a 
conductor. This result suggests an interesting experiment by using two 
small conductors which couple capacitively, and measure the AC response 
as a function of the amplitude of the AC bias. 

\noindent
{\bf Acknowledgments.} 
We gratefully acknowledge financial support by RGC of SAR Government 
of Hong Kong under grant number HKU 7112/97P; by NSF of China;
NSERC of Canada, and FCAR of Qu\'ebec.


\begin{thebibliography}{00}
\bibitem{foot1}
We emphasize that the two-index partial density of states 
$dn_{\alpha\beta}/d\epsilon$ of the scattering matrix theory
\cite{buttiker1} means a completely different quantity as 
compared with the two-index LDOS introduced here. In this paper, 
multiple indices on LDOS denote the nonlinear order in the analysis.
\end{thebibliography}
\end{document}